\documentclass{ptapap}

\author{Cyril Georgy}[1]
\author{Raphael Hirschi}[2,3]
\author{Sylvia Ekstr\"om}[1]
\affil[1]{Department of Astronomy, University of Geneva,\\Chemin des Maillettes 51, 1290 Versoix, Switzerland}
\affil[2]{Astrophysics group, Keele University, Keele, Staffordshire ST5 5BG, UK}
\affil[3]{Kavli Institute for the Physics and Mathematics of the Universe (WPI), Tokyo Institutes for Advanced Study, The University of Tokyo, 5-1-5 Kashiwanoha, Kashiwa, Chiba 277-8583, Japan}

\title{Massive Star Evolution: \\What we do (not) know}

\begin{document}

\maketitle

\begin{abstract}
The modelling of massive star evolution is a complex task, and is very sensitive to the way physical processes (such as convection, rotation, mass loss, etc.) are included in stellar evolution code. Moreover, the very high observed fraction of binary systems among massive stars makes the comparison with observations difficult. In this paper, we focus on discussing the uncertainties linked to the modelling of convection and rotation in single massive stars.
\end{abstract}

\section{Introduction}

The modelling of massive star is a complex task, involving a variety of physical processes. Among the required ingredients of all stellar evolution codes are the treatment of the heat transfer in convective and radiative zones, the nuclear reaction network, the equation of state, the computation of opacities, and the inclusion of mass loss \citep[e.g.][]{Kippenhahn1990a,Maeder2009a}. During the past two decades, a variety of other processes were progressively added, such as rotation \citep{Endal1976a,Zahn1992a,Maeder1998a}, transport of angular momentum and chemical species by internal magnetic fields \citep{Spruit2002a,Maeder2003b} or by internal waves \citep[e.g.][]{Kumar1997a,Talon2002a,Talon2003b,Fuller2014a}. Moreover, massive stars are often found in multiple systems \citep[e.g.][]{Sana2012a}, and their modelling requires in addition the treatment of tidal interactions, Roche-lobe overflow, common envelope evolution and merging \citep[see the review by][]{Langer2012a}.

Each of these processes suffers from uncertainties in the way they should be implemented in stellar evolution code. It leads to major uncertainties in our understanding of the evolution of massive stars, particularly of the post-main-sequence evolution \citep{Martins2013a,Chieffi2013a,Georgy2014a}. In this paper, we focus on discussing some of the uncertainties linked to the modelling of convection and rotation in single massive stars.

\section{The modelling of convection in the interior of massive stars}

\begin{figure}[t]
\includegraphics[width=.48\textwidth]{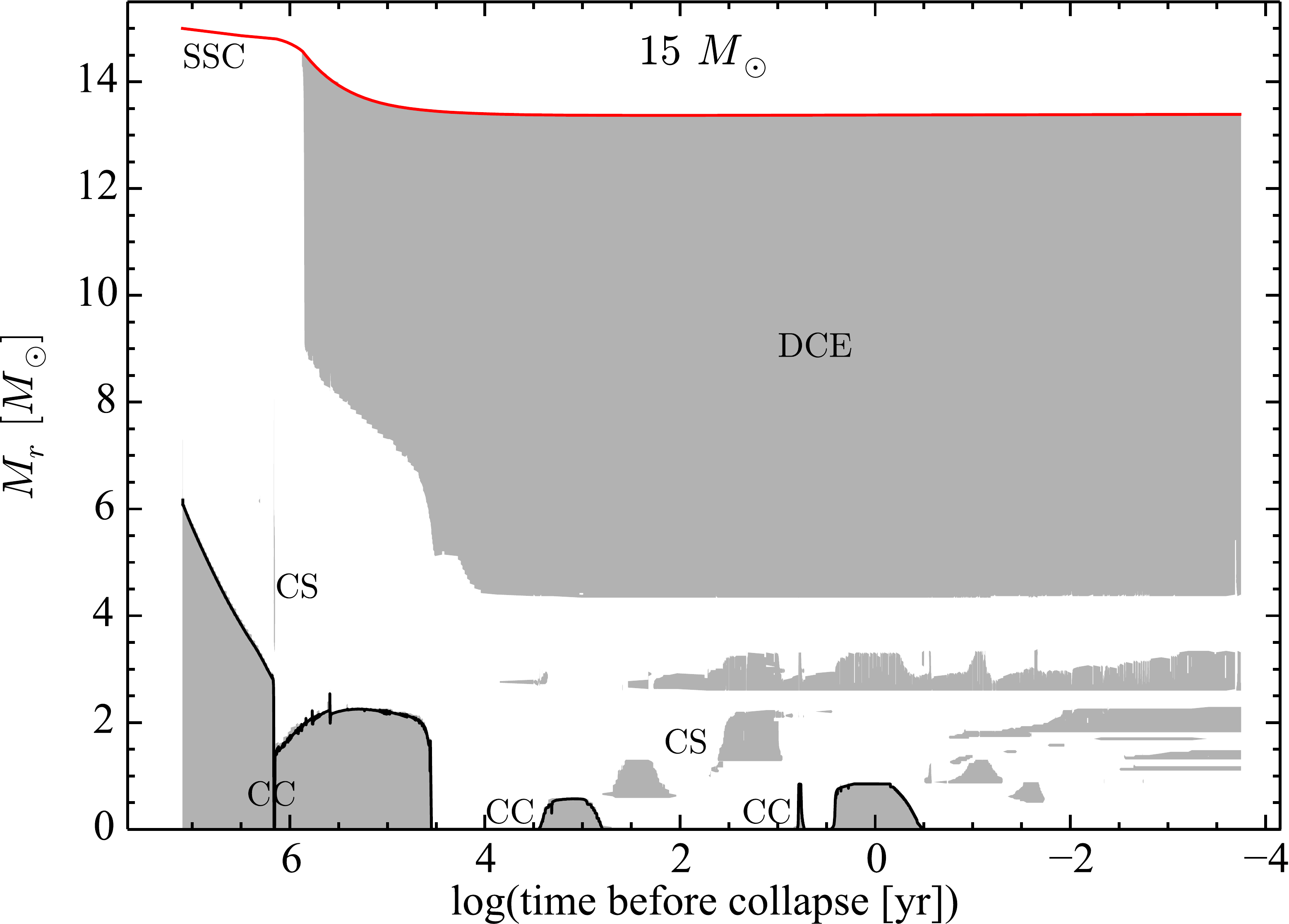}\hfill\includegraphics[width=.48\textwidth]{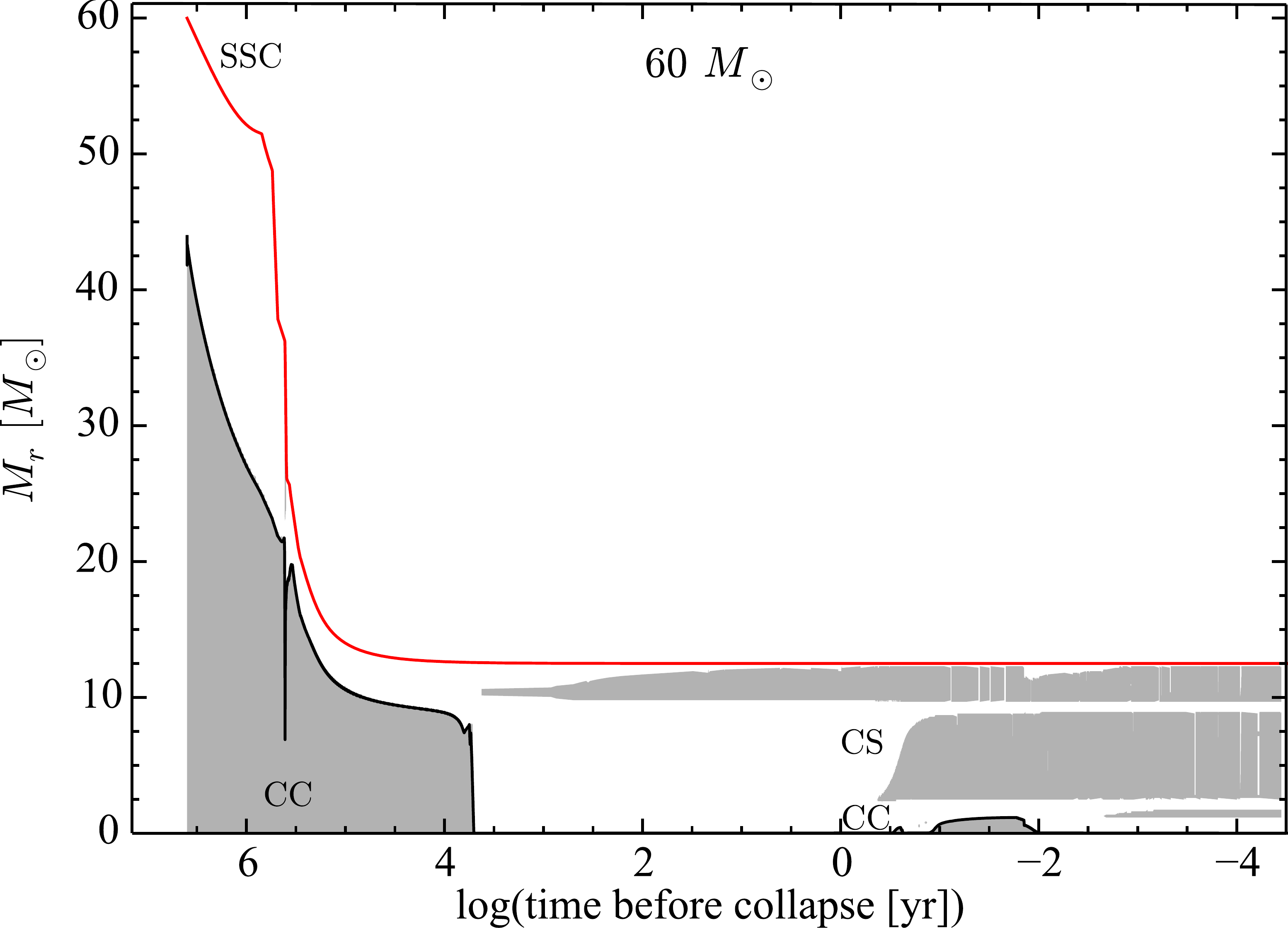}
\caption{Convective structure of solar metallicity $15\,M_\odot$ (left) and $60\,M_\odot$ stellar models (right). The horizontal axis shows the time left until the star collapses. The vertical axis is the mass coordinate (so-called ``Kippenhahn diagram''). The shaded area represent the convective zones inside the star. The labels are for sub-surface convection (``SSC''), convective core (``CC'', highlighted by the thick solid black line), deep convective envelope (``DCE''), and convective shell (``CS''). The thick line indicates the surface of the star.}
\label{fig:Kippen}
\end{figure}

Convection is ubiquitous in massive star evolution (see Fig.~\ref{fig:Kippen}): successive convective cores (``CC'') are often linked to one of the burning stages. After the main-sequence (MS), nuclear burning may also occur in convective shells (``CS''), producing a complex structure during the very late stages of the evolution. During the MS, a very tiny convective zone is present near the surface \citep{Maeder2008b,Cantiello2009a}. Finally, for stars evolving in the red part of the Hertzsprung-Russell diagram (HRD, typically the stars having a red supergiant phase), a deep external convective envelope (``DCE'') develops. This highlights the need for a correct modelling of convection in massive star models.

In most of stellar evolution codes, convection is modelled in two steps:
\begin{enumerate}
\item The position of the convective zone boundaries are determined according to the Schwarzschild or Ledoux criterion \citep[see e.g.][]{Kippenhahn1990a}.
\item Inside the convective zone, the thermal structure is computed by computing a thermal gradient. This can be done in several ways: by assuming that convection is purely adiabatic, by using the ``Mixing-Length Theory'' \citep[MLT][]{Boehm-Vitense1958a} or more sophisticated models \citep[see][and references therein]{Viallet2015a}.
\end{enumerate}

The sizes of the convective cores obtained in this way are known to be too small with respect to observations for a long time \citep[e.g.][]{Maeder1975a}, and need to be artificially extended by an arbitrary length: the ``overshoot''. This overshoot cannot be predicted by design in the framework of the MLT. In stellar evolution codes, the overshoot is usually considered to be ``penetrative'' \citep[][where the core is extended by a constant fraction of the pressure scale height at the edge of the core]{Zahn1991a}, or ``diffusive'', with a diffusion coefficient calibrated on observations or numerical simulations \citep{Freytag1996a,Herwig2000a}.

Both approaches contain free parameters, and thus need to be calibrated. This can be done in different ways, for example by reproducing the observed width of the MS \citep{Ekstrom2012a}, or by fitting the drop of the surface velocities of stars when their surface gravity decreases \citep{Brott2011a}.

On the other hand, the development of multi-dimension hydrodynamics codes and of computing power has allowed for simulations of convection in a variety of physical conditions from first principles: envelope of cool stars \citep[e.g.][]{Freytag2008a,Chiavassa2009a,Viallet2013a,Magic2013a}, or deep convection during different evolutionary stages of star life \citep[e.g.][]{Meakin2007a,Couch2015a,Woodward2015a,Cristini2016b,Muller2016a,Jones2016a}.

Simulations of deep convection show that, at least during the advanced stages, convective boundaries are moving \citep[``entrainment'', see][]{Meakin2007a}.  Moreover, there is a significant mixing across the boundary, making it less stiff than usually accounted for in 1d stellar evolution modelling \citep{Cristini2016b}.

From observations, a similar result is obtained thanks to asteroseismology for the overshoot during the MS \citep[see e.g.][]{Moravveji2016a}. The modelling of variable blue supergiants seems to be a promising way of constraining convection is stellar models \citep{Georgy2014a}. In any case, it is clear that MLT is not able to correctly reproduce the behaviour of convective flows as seen in multi-d simulations, and there is an urgent need for a new way of treating convection in stellar evolution models \citep[see][]{Arnett2015a}.

\section{The modelling of rotation in the interior of massive stars}

The inclusion of rotation in 1d stellar evolution codes is not straightforward. Due to its (at least) 2d nature, several hypotheses and approximations are requested to treat rotation in 1d (process sometimes called ``1.5d''). First of all, the star is described in the framework of the Roche model, assuming a spherical-symmetry gravitational potential on top of which the effects of centrifugal acceleration are added. Moreover, a strong horizontal turbulence is assumed inside the star, homogenising the angular velocity on an isobar \citep[``shellular rotation'', see][]{Zahn1992a,Maeder1998a}.

In this framework, rotation has two main effects:
\begin{enumerate}
\item the centrifugal acceleration modifies the usual stellar structure equations by adding corrective terms in the momentum equation and radiative transfer equation \citep{Meynet1997a}.
\item thermal non-equilibrium produces large scale currents inside the star \citep[the so-called meridional circulation, or Eddington-Sweet currents,][]{Sweet1950a}. In turn, these currents transport angular momentum and chemical species, modifying the internal rotation of the star. Differential rotation can occur, generating shear turbulence.
\end{enumerate}

The transport of angular momentum is modelled by the following relation \citep{Zahn1992a,Maeder1998a}:
\begin{equation}
\rho\partial_t\left(r^2\bar{\Omega}\right) = \frac{1}{5r^2}\partial_r\left(\rho r^4\bar{\Omega}U_2\right) + \frac{1}{r^2}\partial_r\left(\rho D_v r^4\partial_r\bar{\Omega}\right),\label{eq:advection}
\end{equation}
where $\bar{\Omega}$ is the mean angular velocity on an isobar, $U_2$ is the radial component of the meridional circulation velocity, and $D_v$ is the vertical turbulence diffusion coefficient. The expression of $U_2$ is complex and can be found in \citet{Maeder2009a}.

The transport of chemical species can be modelled by a purely diffusive approach \citep{Chaboyer1992a}:
\begin{equation}
\rho\partial_t\left(X_i\right) = \frac{1}{r^2}\partial_r\left(\rho r^2\left(D_v + D_\mathrm{eff}\right)\partial_r\left(X_i\right)\right),
\end{equation}
where $X_i$ is the abundance of the element $i$, and $D_\mathrm{eff}$ is the effective diffusion coefficient, accounting for the effect of meridional circulation: $D_\mathrm{eff} = \frac{\left(rU_2\right)^2}{30D_\mathrm{h}}$, where $D_\mathrm{h}$ is the horizontal turbulence diffusion coefficient.

\subsection{The vertical turbulence}
The diffusion coefficient $D_v$ should account for any kind of turbulence arising in the vertical direction. Depending on which stellar evolution code is used, different effects are accounted for: secular shear instability \citep{Maeder1997a,Talon1997a}, dynamical shear instability \citep[e.g.][]{Heger2000a}, Solberg-Hoiland instability \citep[e.g.][]{Heger2000a}, Goldreich-Schubert-Fricke instability \citep[e.g.][]{Heger2000a, Hirschi2010b}, Tayler-Spruit dynamo induced mixing \citep{Spruit2002a,Maeder2003b}. Most of time, the corresponding diffusion coefficients are summed up. However, \citet{Maeder2013a} propose a way to consider the combined effects of these instabilities at once.

\subsection{The horizontal turbulence}
As in the case for the vertical turbulence, several prescriptions can be found in the literature for the diffusion coefficient linked to the horizontal turbulence \citep{Zahn1992a,Maeder2003a,Mathis2004a}.

\subsection{Advection and diffusion or diffusion only?}

As of today, we can distinguish two big families among the stellar evolution codes that contains the treatment of stellar rotation. The first of them solves the full equation for the transport of angular momentum (eq.~\ref{eq:advection}): the Geneva stellar evolution code \citep{Eggenberger2008a}, STAREVOL \citep{Decressin2009b}, FRANEC \citep{Chieffi2013a}, ROSE \citep{Potter2012a}\footnote{\footnotesize{This list is indicative only and has no ambition to be complete. It mostly covers the codes used in the massive star community.}}. On the other hand, other codes uses an approximate form of eq.~\ref{eq:advection}, where the advective term is replaced by another diffusion coefficient, making this equation fully diffusive: MESA  \citep{Paxton2013a}, STERN \citep{Petrovic2005a}, or Kepler \citep{Heger2005a}. There is so far no consensus about which implementation should be used. However, the reader should keep in mind that both implementations provide different results in terms of evolution of the surface velocities and chemical species \citep{Ekstrom2012a,Georgy2013a,Brott2011a,Chieffi2013a,Martins2013a}.

Another caution is linked to the uncertainty of the choice of the vertical or horizontal diffusion coefficients. Different choices can also lead to qualitatively different results \citep{Meynet2013a}.

\subsection{Effects of rotation on stellar evolution}

\paragraph{Surface abundances.} One of the most important effect of rotation is the modification of the surface chemical composition as a function of time, due to the rotation-induced internal mixing. For most of massive star models, it implies that chemical species produced in the core of the star during hydrogen-burning are progressively brought to the surface. For example, it implies an increase of the N abundance, while C and O abundances decrease. This effect is generally stronger for higher mass star and at lower metallicity \citep{Ekstrom2012a,Georgy2013a}. In the most extreme cases, internal mixing favours the evolution towards the Wolf-Rayet stage, making it occur earlier in the lifetime of a massive star, or for lower initial mass stars \citep[e.g.][]{Georgy2012b}.

\paragraph{Tracks in the HRD.} On the Zero-Age Main-Sequence, a rotating model is cooler and less luminous than its non-rotating counterpart. This is due to the support of the centrifugal acceleration, making the model behaves as a lower mass one. After the ZAMS, internal mixing brings fresh hydrogen into the core, making its mass diminish more slowly, and thus keeping the star at higher luminosity. At the same time, the change in the surface abundances (more helium, less hydrogen) makes the star evolve at higher effective temperature than in the non-rotating case \citep{Meynet2000a}. In some extreme cases, for very rapidly rotating star, the mixing is so efficient that the star can evolve nearly homogeneously \citep{Yoon2005a,Meynet2007a,Szecsi2015a}.

\paragraph{Lifetimes.} Due to the ingestion of fresh hydrogen by the core during the MS due to rotational mixing, the lifetime of the star are increased \citep{Georgy2013a}.
The increase of the lifetime can reach several tens of percent with respect to the non-rotating case. This has consequence on the computation of isochrones \citep{Georgy2014b}.

\section{Conclusions}

In this paper, we have briefly discussed the implementation of convection and rotation in stellar evolution codes, in particular in the context of massive star evolution. Current implementations of convection are described, and we highlight the shortcomings shown by recent multi-dimensional hydrodynamics simulations or observations. The various ways of dealing with rotation are also explained. Finally, we discussed some impacts of the inclusion of rotation on our understanding of stellar evolution.

\acknowledgements{CG and SE acknowledge support from the Swiss National Science Foundation (project number 200020-160119). RH acknowledges support from the European Research Council under the European Union's Seventh Framework Programme (FP/2007-2013) / ERC Grant Agreement n. 306901 and WPI IPMU.}

\bibliographystyle{ptapap}
\bibliography{Georgy_bib}

\end{document}